\documentclass[prb,aps,nobalancelastpage,showpacs,amssymb,twocolumn,citeautoscript]{revtex4-1}

\usepackage{graphicx}

\begin{document}
\preprint{PRB/Y. K. Li et al.}

\title{Superconductivity induced by La doping in Sr$_{1-x}$La$_{x}$FBiS$_{2}$ system}

\author{Xi Lin$^{1}$, Xinxin Ni$^{1}$, Bin Chen$^{1}$, Xiaofeng Xu$^{1}$, Xuxin Yang$^{1}$, Jianhui Dai$^{1}$, Yuke Li$^{1,2,*}$\footnote[1]{Electronic address: yklee@hznu.edu.cn}, Xiaojun Yang$^{2}$, Yongkang Luo$^{2}$, Qian Tao$^{2}$, Guanghan Cao$^{2}$ and Zhuan Xu$^{2,*}$\footnote[2]{Electronic address: zhuan@zju.edu.cn}}

\affiliation{$^{1}$Department of Physics, Hangzhou Normal University, Hangzhou 310036, China\\
$^{2}$State Key Lab of Silicon Materials and Department of Physics, Zhejiang University, Hangzhou 310027, China\\}

\date{\today}

\begin{abstract}

Through a combination of X-ray diffraction, electrical transport, magnetic susceptibility, and heat capacity measurements, we report the effect of La
doping on Sr in the newly discovered SrFBiS$_{2}$ system. Superconducting transition with critical temperature \emph{T$_{c}$} of 2.8 K, developed from
a semiconducting-like normal state, was found in Sr$_{0.5}$La$_{0.5}$FBiS$_{2}$. A strong diamagnetic signal and a clear specific heat anomaly
associated with this transition were observed, confirming bulk superconductivity. The upper critical field $H_{c2}(0)$ was
estimated to be $~$1 Tesla by using the Ginzburg-Landau approach. Our experiments therefore demonstrate that bulk superconductivity can be achieved by electron doping in the SrFBiS$_{2}$ system.

\end{abstract}

\maketitle

Superconductivity within layered crystal structure has attracted sustained interest in the community of correlated electron systems, primarily due to
the exotic superconducting properties they exhibit. Notable examples include the high-\emph{T$_{c}$} cuprates with the CuO$_{2}$ plane and the
Fe-based superconductors with Fe$_{2}$\emph{An}$_{2}$ (\emph{An} = P, As, Se) superconducting(SC) layers. Recently, superconductivity in the novel
BiS$_{2}$-based superconductor Bi$_{4}$O$_{4}$S$_{3}$ with a superconducting transition temperature (\emph{T$_{c}$}) of 8.6 K has been
reported\cite{BOS}. Immediately following this result, several \emph{Ln}O$_{1-x}$F$_{x}$BiS$_{2}$(\emph{Ln}=La, Ce, Pr, Nd) superconductors, with the
highest \emph{T$_{c}$} of $~$10 K\cite{LaFS,NdFS,LaFS2,CeFS,PrFS}, have been then discovered. While \emph{T$_{c}$} is largely determined by
[Ln$_{2}$O$_{2}$]$^{2-}$ blocking layer in these systems, all these superconductors share the common BiS$_{2}$ layer. Although these BiS$_{2}$-based
layered superconductors show a wealth of similarities to the iron pnictides, some differences are also manifest. First, the parent compound
\emph{Ln}OBiS$_{2}$ is a bad metal without detectable antiferrimagnetic transition or structure phase transition, implying that magnetism
is of less importance to superconductivity in BiS$_{2}$-based systems. Second, resistivity measurements in BiS$_{2}$-based systems suggest that
superconductivity appears in close proximity to an insulating normal state for the optimal superconducting sample\cite{CeFS}, in sharp contrast to
the iron based superconductors where superconductivity grows from a metallic state. Interestingly, pairing symmetry in layered BiS$_{2}$-based
compounds has yet to be determined albeit some theorists conjectured that BiS$_{2}$ based compounds are strongly electron-phonon coupled
superconductors, and the pairing symmetry may be an extended $s$-wave with full gaps on different parts of Fermi surfaces\cite{Yildirim,HuJP}, in
analogy to the iron-based superconductors\cite{Mazin,Kuroki}. However, the experimental evidence of pairing symmetry is still lacking thus far.

Very recently, a new BiS$_{2}$ based layered compound SrFBiS$_{2}$ has been reported\cite{SrF}. This compound is isostructural to LaOBiS$_{2}$, with the [Ln$_{2}$O$_{2}$]$^{2-}$ layer being replaced by iso-charged [Sr$_{2}$F$_{2}$]$^{2-}$ block, similar to the case of
LaOFeAs and SrFFeAs. The resistivity of SrFBiS$_{2}$ shows the semiconducting behavior, however, no
superconductivity has been reported in this system.

In the paper, we report the successful synthesis and the detailed characterization of La-doped Sr$_{1-x}$La$_{x}$FBiS$_{2}$ ($x$ = 0, 0.5) samples.
The parent compound SrFBiS$_{2}$ exhibits thermally excited resistivity down to low temperature, consistent with the existing
literature\cite{SrF}. As $x$ = 0.5, resistivity at normal state still shows the semiconducting behavior, but its absolute value drops by a factor of 2, followed by a sharp superconducting transition below $~$2.8 K. The strong diamagnetic signal and a clear specific heat jump provide compelling
evidence of bulk superconductivity in Sr$_{1-x}$La$_{x}$FBiS$_{2}$ system. Finally, the specific heat data show evidence of
strong-coupling nature of the observed superconductivity.

The polycrystalline samples of Sr$_{1-x}$La$_{x}$FBiS$_{2}$ ($x$ = 0, 0.5) used in this study were synthesized by two-step solid state reaction method. The starting materials, La pieces, SrF$_{2}$ powder, Bi and S powder are all of high purity ($\geq$99.9\%). Firstly, La$_{2}$S$_{3}$ was
pre-synthesized by reacting stoichiometric S powder and La pieces at 873 K for 10 hours. After that, the as-grown La$_{2}$S$_{3}$ and the powder of
SrS, SrF$_{2}$, Bi, and S as starting material were weighted according to their stoichiometric ratio and then fully ground in an agate mortar. The
mixture of powder was then pressed into pellets, heated in an evacuated quartz tube at 1073 K for 10 hours and finally furnace-cooled to room
temperature. The whole process was repeated once again in order to get the pure phase.

Crystal structure characterization was performed by powder X-ray diffraction (XRD) at room temperature using a D/Max-rA diffractometer with Cu
K$_{\alpha}$ radiation and a graphite monochromator. Lattice parameters were obtained by Rietveld refinements. The electrical resistivity was measured with a standard four-terminal method covering temperature range from 0.4 to 300 K in a commercial
Quantum Design PPMS-9 system with a $^{3}$He refrigeration insert. Specific heat and Hall effect measurements were also performed in this system. The temperature dependence of d.c. magnetization was measured on a Quantum Design MPMS-5.

\begin{figure}[h]
\includegraphics[width=8cm]{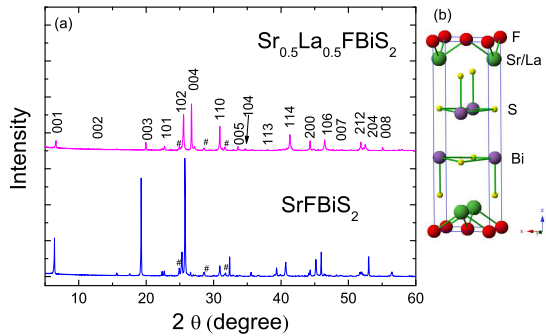}
\caption{(Color online) (a) Powder X-ray diffraction patterns of Sr$_{1-x}$La$_{x}$FBiS$_{2}$ ($x$ $=$ 0, 0.5) samples. The \# peak positions designate
the impurity phase of Bi$_{2}$S$_{3}$. (b) Crystal structure of Sr$_{1-x}$La$_{x}$FBiS$_{2}$. The solid line indicates the unit cell.}
\end{figure}

Figure 1 (a) shows the powder XRD patterns of the Sr$_{1-x}$La$_{x}$FBiS$_{2}$ ($x$=0, 0.5) samples. Overall, the main diffraction peaks of these two
samples can be well indexed based on a tetragonal cell structure with the P4/nmm space group. Extra minor peaks arising from impurity phase of
Bi$_{2}$S$_{3}$ can also be seen. The content of impurity phase Bi$_{2}$S$_{3}$
estimated by Rietveld fitting is about 15\% and 12\% for parent compound($x = 0$) and $x = 0.5$ sample, respectively. The lattice parameters of the parent compound SrFBiS$_{2}$ are extracted to be $a=$ 4.0820{\AA} and $c=$ 13.8025{\AA} by Rietveld fitting, in good agreement with the previous report\cite{SrF}. For the La-doped sample, however, the (00$l$)
peaks shift discernibly towards higher 2$\theta$ angles, suggesting the smaller $c$-axis constant. Thus, while the $a$-axis lattice parameter
increases slightly to 4.0822{\AA} in the La-doped sample, the $c$-axis decreases significantly to 13.335{\AA}, implying that the La impurity was
indeed doped into the lattice, given that the ionic radius of La ion is known to be smaller than that of Sr. This is quite similar to the case of
LaO$_{1-x}$F$_{x}$BiS$_{2}$\cite{LaFS2}. The resultant crystal structure is therefore schematically shown in the Fig.1(b), which is composed of
stacked rock-salt-type BiS$_{2}$ layer alternating with fluorite-type SrF layer along the $c$-axis.

The zero-field temperature dependence of the electrical resistivity ($\rho$) of these two samples is plotted in Fig. 2(a). For the parent
SrFBiS$_{2}$ compound, the resistivity clearly shows thermally activated behavior with decreasing temperature from 300 K, and no resistivity anomaly can be observed down to 2 K. The thermal activation energy ($E_a$) obtained by fitting with the thermal activation formula $\rho(T)=\rho_0 \exp(E_a/k_B T)$ for the temperature range from 100 K to 300 K is about 38.2 meV, consistent with previous studies\cite{SrF}. However, it is noted that its absolute value is nearly two thirds smaller than the reported value by Lei et al\cite{SrF}. We attribute this to the impurity phase of Bi$_{2}$S$_{3}$ with sulfur deficiency\cite{BiS} in our sample which may decrease the absolute value of resistivity. As Sr is partially replaced by La ($x$ = 0.5), the resistivity remains semiconducting-like at high temperatures,
with the magnitude of resistivity substantially reduced compared to the undoped sample. Meanwhile, the $E_a$ fitted at high temperature also decreases to 8.6 meV, suggesting the decrease of gap size because of electron doping. With further cooling down, a sharp superconducting transition
with \emph{T$_{c}$} of 2.8 K is clearly seen. This result is reminiscent of LaO$_{0.5}$F$_{0.5}$BiS$_{2}$, where the normal state shows
semiconducting behavior, and it undergoes superconducting transition below $~$10 K\cite{LaFS}. Interestingly, the first principles calculations\cite{Yildirim} has suggested that there may be a charge density wave instability or an enhanced correlation
effect in this system. Yet, no such instability can be detected merely from the resistivity measurement given in this study.

\begin{figure}[h]
\includegraphics[width=8cm]{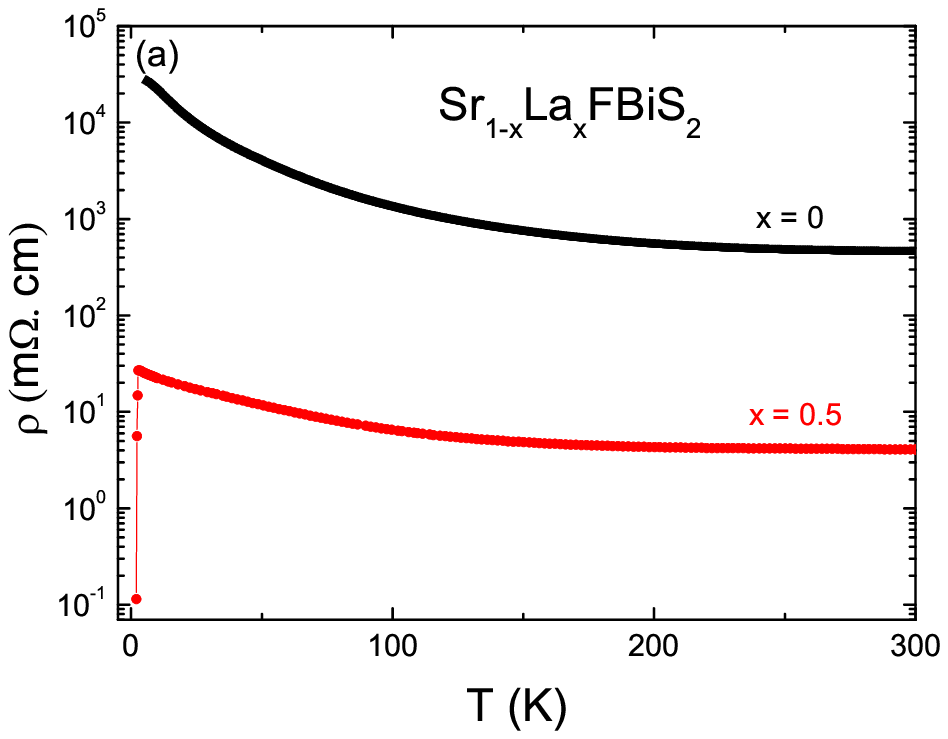}
\includegraphics[width=8cm]{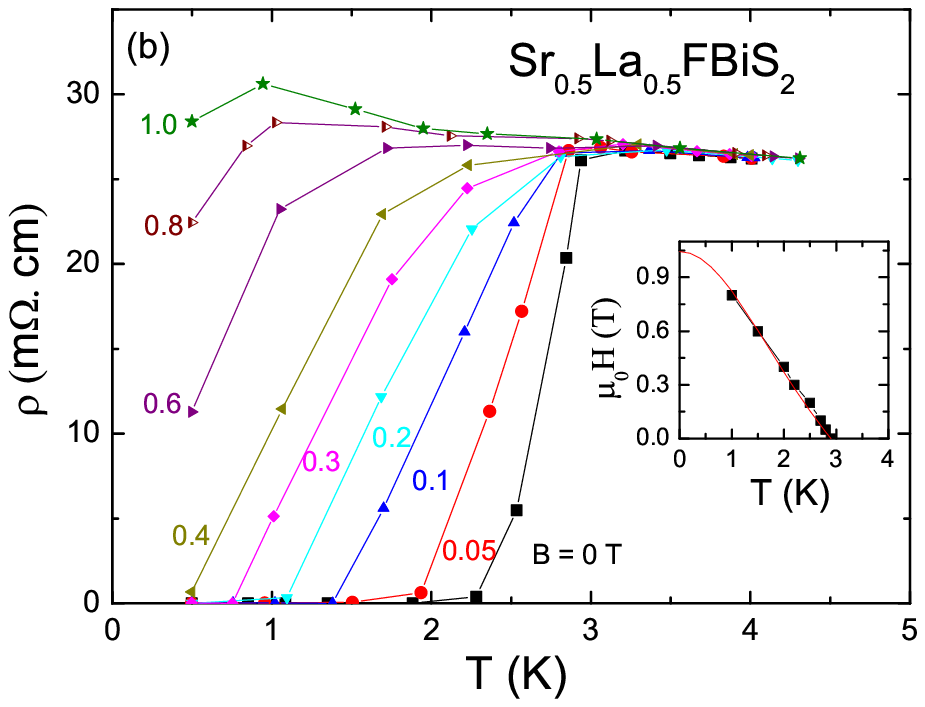}
\caption{(Color online)(a)Temperature dependence of resistivity ($\rho$) for the Sr$_{1-x}$La$_{x}$FBiS$_{2}$ ($x$ $=$ 0, 0.5) samples. (b)Temperature
dependence of resistivity for Sr$_{0.5}$La$_{0.5}$FBiS$_{2}$ samples under several different magnetic field (B $=$ 0, 0.05, 0.1, 0.2, 0.3, 0.4, 0,6, 0.8, 1 T).}
\end{figure}
Figure 2 (b) shows the close-up view of the low-temperature resistivity for Sr$_{0.5}$La$_{0.5}$FBiS$_{2}$ samples under various magnetic fields
below 4 K. The inset gives the temperature dependence of the upper critical field $\mu_0$$H_{c2}(T)$, determined by using 90\% normal state
resistivity criterion. The temperature dependence of $\mu_{0}H_{c2}(T)$ is nearly linear in the temperature range studied. According to
Ginzburg-Landau theory, the upper critical field $H_{c2}$ evolves with temperature following the formula:
\begin{equation}
{H_{c2}(T) = H_{c2}(0)(1-t^2)/(1+t^2)},
\end{equation}
where \emph{t} is the renormalized temperature \emph{T/T$_{c}$}. It is found that the upper critical field $H_{c2}$ as drawn can be well fitted by
this model and its zero temperature limit is estimated to be 1.04 T.

In order to verify the bulk nature of the observed superconductivity, the temperature dependence of d.c. magnetic susceptibility was measured under 5
Oe magnetic field with both zero field cooling (ZFC) and field cooling (FC) modes for the superconducting samples. The strong diamagnetic signal was
observed for Sr$_{0.5}$La$_{0.5}$FBiS$_{2}$ sample. The \emph{T$_{c}$} value determined from the magnetic susceptibility agrees well with the one
extracted from the resistivity data. The estimated volume fraction of magnetic shielding from ZFC data is over 30\%, bearing out the bulk
superconductivity. Note that the Meissner volume fraction estimated by FC measurement data also exceeds 20\%, indicating the high quality
superconducting sample.

\begin{figure}
\includegraphics[width=8cm]{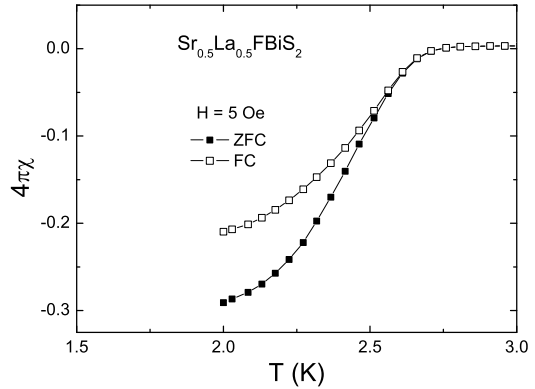}
\caption{(Color online) The diamagnetic signal vs. Temperature under a magnetic field of 5 Oe with ZFC (Solid) and FC(Open) modes for
Sr$_{0.5}$La$_{0.5}$FBiS$_{2}$}
\end{figure}

To further study its superconducting properties, the specific heat measurement of
Sr$_{0.5}$La$_{0.5}$FBiS$_{2}$ sample was performed in the temperature range from 0.5 K to 5 K. As
plotted in Fig. 4, a clear specific-heat jump can be observed at \emph{T$_{c}$} = 2.8 K, further
confirming the bulk superconductivity. Below 1.5 K, however, the low temperature heat capacity
undergoes a moderate upturn which very likely arises from Schottky anomaly associated with the
impurities phase or the freezing out of La nuclear spins\cite{Cava}. Taking into account the
Schottky anomaly component, the overall specific heat can be written as:
\begin{equation}
{C/T = \gamma + \beta T^2 + \alpha/T^3}
\end{equation}
where $\gamma$ and $\beta$\emph{T}$^2$ account for the electronic and lattice contributions respectively, and the last term, $\alpha/T^3$, is the
Schottky anomaly. The plot of \emph{C/T}-$\alpha/T^3$ versus \emph{T}$^{2}$ shows a linear behavior in the high temperature range from 2.9 to 5 K
(data not explicitly shown here). The resultant electronic coefficient $\gamma$ and the lattice coefficient $\beta$ are 1.42 mJ/ mol K$^{2}$ and
$\beta$ = 0.414 mJ/ mol K$^{4}$ respectively. The Debye temperature is then estimated to be 265 K. This value is somewhat smaller that those of
LaONiAs\cite{LaNiFe}, LaOFeP\cite{LaFeP} and LaO$_{0.89}$F$_{0.11}$FeAs\cite{LaFAs}. The lattice contribution was further subtracted in
the inset in order to see the clear specific heat jump at the SC transition. The corresponding electronic specific jump at \emph{T$_{c}$}, therefore,
is estimated to be $\Delta$ C$_{e}$/$\gamma$\emph{T$_{c}$} $=$ 1.4, very close to the weak-coupling Bardeen-Cooper-Schrieffer (BCS) value of 1.43.
Considering the relatively low volume fraction of magnetic shielding suggested from susceptibility measurement, the real size of $\Delta$
C$_{e}$/$\gamma$\emph{T$_{c}$} ought to be much larger than the BCS value, implying that SC in Sr$_{0.5}$La$_{0.5}$FBiS$_{2}$ likely involves a
strong electron-phonon coupling, consistent with the theoretical prediction\cite{Yildirim}.

\begin{figure}[h]
\includegraphics[width=8cm]{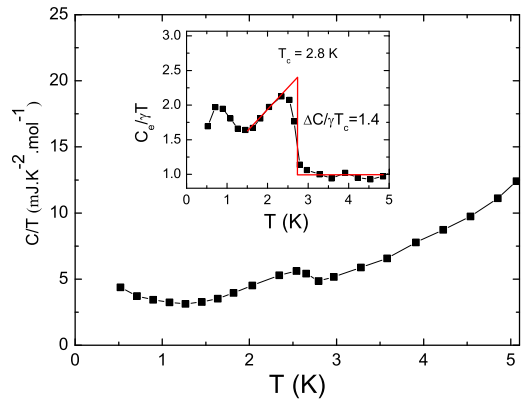}
\caption{(Color online) The specific heat of Sr$_{0.5}$La$_{0.5}$FBiS$_{2}$ samples under zero field below 5 K. The inset shows $\Delta$
C$_{e}$/$\gamma$\emph{T$_{c}$} as a function of temperature, where C$_{e}$ is the electronic specific heat.}
\end{figure}

Finally, the temperature dependence of Hall coefficient $R_{H}$ for Sr$_{0.5}$La$_{0.5}$FBiS$_{2}$ sample is shown in Fig. 5. The negative $R_{H}$
signal covers the whole temperature region, indicating that electron-type charge carrier is dominant in this sample. Clearly, $R_{H}$ is
$T$-independent at high temperature regime, and drops drastically below 50 K. The result reminds us of the similar behavior observed in iron-based
SC, where $R_{H}$ decreases sharply due to spin density wave(SDW) formation\cite{Oak}. In the present sample, however, no SDW/CDW anomaly has been observed in the measurements of resistivity and magnetization. On the other hand, it is known that the inherent impurity phase may act to smear out the anomalies in the resistivity and susceptibility. In this regard, it would be interesting to clarify this problem by first principles calculations\cite{Yildirim} or further improving  the sample quality. In a single band metal, the Hall coefficient $R_{H}$ is associated with carrier density (\emph{n}) as $R_{H}$ =
1/\emph{ne}. We therefore estimate the charge carrier density to be 1.05$\times$10$^{19}$/cm$^{3}$ at 5 K, which is similar to the
Bi$_{4}$O$_{4}$S$_{3}$ system\cite{WHHBOS} and iron based superconductors\cite{Oak2} with a very low superfluid density. In addition, we can not rule out the multi-band effect\cite{CeFS}, as the origin of the drop in $R_{H}$.

\begin{figure}
\includegraphics[width=8cm]{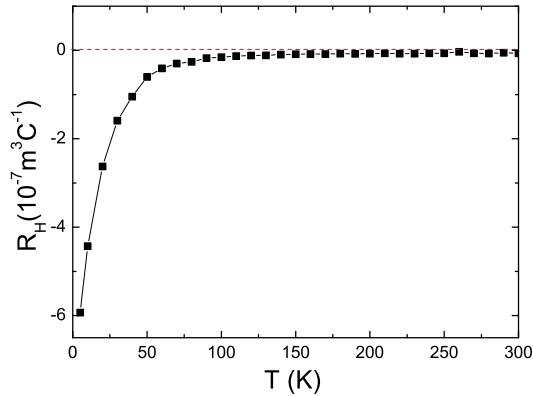}
\caption{(Color online) Temperature dependence of Hall coefficient $R_{H}$ measured at 5 T for Sr$_{0.5}$La$_{0.5}$FBiS$_{2}$ sample.}
\end{figure}

In summary, we study the electron doping effect, i.e., partially substituting La for Sr, in the newly discovered SrFBiS$_{2}$ system. The parent
compound shows semiconducting-like resistivity with decreasing temperature. Via electron doping, we discovered a new BiS$_{2}$-based superconductor
Sr$_{0.5}$La$_{0.5}$FBiS$_{2}$ with a \emph{T$_{c}$} as high as 2.8 K. Importantly, this superconductivity was found to develop on the background of
insulating normal state, raising a question of whether this superconductor is in proximity to a Mott insulator or a band insulator. Based on the heat
capacity measurement, we speculate that the superconducting electron pairs might have a strong-coupled nature. In this regard, it would certainly
be interesting to identify the symmetry of superconducting order parameter in this system along with other BiS$_{2}$-based superconductors.

We would like to thank Chao Cao for discussions. This work is supported by the National Basic Research Program of China (Grant No. 2011CBA00103 and 2012CB821404), NSFC (Grant No. 11174247, 11104053 and 11104051).


\begin{thebibliography}{00}


\bibitem{BOS} Y. Mizuguchi, H. Fujihisa, Y. Gotoh, K. Suzuki, H. Usui, K. Kuroki, S. Demura, Y. Takano, H. Izawa, O. Miura, arXiv: 1207.3145
\bibitem{LaFS} Y. Mizuguchi, S. Demura, K. Deguchi, Y. Takano, H. Fujihisa, Y. Gotoh, H. Izawa, O. Miura, J. Phys. Soc. Jap. \textbf{81}  114725 (2012)
\bibitem{NdFS}S. Demura, Y. Mizuguchi, K. Deguchi, H. Okazaki, H. Hara, T. Watanabe, S. J. Denholme, M. Fujioka, T. Ozaki, H. Fujihisa, Y. Gotoh, O. Miura, T. Yamaguchi, H. Takeya, and Y. Takano, arXiv: 1207.5248
\bibitem{LaFS2}V.P.S. Awana, A. Kumar, R. Jha, S. Kumar, J. Kumar, and A. Pal, arXiv: 1207.6845
\bibitem{CeFS} J. Xing, S. Li, X. Ding, H. Yang and H. H. Wen, arXiv: 1208.5000
\bibitem{PrFS} R. Jha, S. K. Singh, and V. P. S. Awana, arXiv: 1208.5873
\bibitem{Yildirim} T. Yildirim, arXiv: 1210.2418
\bibitem{HuJP} Y. Liang, X. Wu, W. F. Tsai, and J. P. Hu, arXiv: 1211.5435

\bibitem{Mazin} I. I. Mazin, D. J. Singh, M. D. Johannes, and M. H. Du, Phys.
Rev. Lett. \textbf{101}, 057003 (2008)
\bibitem{Kuroki} K. Kuroki, S. Onari, R. Arita, H. Usui, Y. Tanaka, H. Kontani, and H. Aoki,
Phys. Rev. Lett. \textbf{101} 087004 (2008).

\bibitem{SrF} H. C. Lei, K. F. Wang, M. Abeykoon, E. S. Bozin, and C. Petrovic, arXiv: 1208.3183

\bibitem{BiS} B. Chen, C. Uher, L. Iordanidis, and M. G. Kanatzidis, Chem. Mater. \textbf{9}, 1655 (1997)



\bibitem{Cava}T. M. McQueen, T. Klimczuk, A. J. Williams, Q. Huang, and R. J. Cava, Phys. Rev. B \textbf{79}, 172502 (2009)
\bibitem{LaNiFe} Z. Li, G. Chen, J. Dong, G. Li, W. Hu, D. Wu, S. Su, P. Zheng, T. Xiang, N. L. Wang, J. L. Luo, Phys. Rev. B \textbf{78}, 060504(R) (2008)
\bibitem{LaFeP} T. M. McQueen, M. Regulacio, A. J. Williams, Q. Huang, J. W. Lynn, Y. S. Hor, D. V. West, M. A. Green, and R. J. Cava, Phys. Rev. B \textbf{78}, 024521 (2008)
\bibitem{LaFAs} Y. Kohama, Y. Kamihara, M. Hirano, H. Kawaji, T. Atake, and H. Hosono, Phys. Rev. B \textbf{78}, 020512(R) (2008).


\bibitem{WHHBOS}  S. Li, H. Yang, J. Tao, X. Ding, and H. H. Wen, arXiv: 1207.4955

\bibitem{Oak}M. A. McGuire, A. D. Christianson, A. S. Sefat, B. C. Sales, M. D. Lumsden, R. Jin, E. A. Payzant, D. Mandrus, Y. Luan, V. Keppens, V. Varadarajan, J. W. Brill, R. P. Hermann, M. T. Sougrati, F. Grandjean, and G. J.
Long, Phys. Rev. B \textbf{78}, 094517 (2008).

\bibitem{Oak2} Athena S. Sefat, Michael A. McGuire, Brian C. Sales, Rongying Jin, Jane Y. Howe, and David Mandrus, Phys. Rev. B \textbf{77}, 174503 (2008)

\end{thebibliography}
\end{document}